# First principles investigation of high thermal conductivity in hexagonal germanium carbide(2H-GeC)

Rajmohan Muthaiah, Jivtesh Garg

School of Aerospace and Mechanical Engineering, University of Oklahoma, Norman, OK-73019, USA

**Abstract:** Designing and searching for a high thermal conductivity material in both bulk and nanoscale is highly demanding for electronics cooling. In this work, we studied the thermal conductivity of 2H-Germanium Carbide(2H-GeC) using first principles calculations. At 300 K, we are reporting a high thermal conductivity of 1350 $Wm^{-1}K^{-1}$ and 1050 $Wm^{-1}K^{-1}$ along a-axis and c-axis respectively for pure 2H-GeC. These values are 130% higher than the thermal conductivity of 2H-silicon carbide and 20% lower than cubic germanium carbide(c-GeC). We analyzed the phonon group velocities, phonon scattering rates and mode contribution from acoustic and optical phonons. We also studied the thermal conductivity of nanostructured 2H-GeC for heat dissipation in nanoelectronics. At room temperature, thermal conductivity of 2H-GeC is ~65 $Wm^{-1}K^{-1}$ at nanometer length scales(L) of 100 nm is equal to that of the c-GeC. This result suggests that, 2H-GeC will be a promising material for thermal management applications in micro/nano electronics.

Modern integrated circuits chips consist of millions of transistors generate heat fluxes in a very small areas knows as hot spots has to be removed to improve the reliability[1-4]. Thermal conductivity in nanostructure differs from those in the bulk systems because of its meanfreepath are comparable to the characteristic length of the nanostructures[5]. Thermal conductivity is reduced significantly when the dimensions are reduced to nanoscale due to the boundary scattering[6, 7]. In bulk materials with phonon mean free path of the acoustic phonons higher than the system size will be scattered due to the boundary scattering when the system is reduced. Hence, we need a material with high thermal conductivity in both bulk and nanoscale. Typically, acoustic phonons carry most of their heat due to its high phonon group velocities and vibrational frequencies and optical phonons being the scattering channels for the acoustic phonons through Umklapp scattering. Various materials were reported with optical phonon frequencies having a significant contribution to its overall thermal conductivity and plays is critical for heat conduction in nanoscale because of its phonon MFP is in the range of few nanometers (< 100 nm)[8-11]. Despite having an ultra-high thermal conductivity($k$) of 1517 Wm$^{-1}$K$^{-1}$ in cubic germanium carbide(c-GeC)[12], $k$ of its wurtzite structure is unknown. Bulk[13] and monolayer[14] hexagonal germanium carbide was reported for electronic properties and there is lack of study for its thermal conductivity. In this work, we are reporting an ultra-high thermal conductivity of 1350 Wm$^{-1}$K$^{-1}$(1050 Wm$^{-1}$K$^{-1}$) along the a-axis(c-axis) in hexagonal germanium carbide(2H-GeC) which is 130% higher than the hexagonal silicon carbide(2H-SiC)[15]. We systematically investigate the contributions from transverse acoustic (TA), longitudinal acoustic (LA) and optical phonons modes. Our first principles calculations reveal that, optical phonons with high phonon group velocity contributes ~46% to its overall thermal conductivity at room temperature (300 K). We also report the length dependence thermal conductivity between 10 nm-1000 nm for its application in nanoscale. At nanometer length scales of L=100 nm, a high thermal conductivity of ~65 Wm$^{-1}$K$^{-1}$. High thermal conductivity of 2H-GeC in both bulk and nanoscale indicates that, 2H-GeC will be a promising material for thermal management applications.

First principles computations were performed using local density approximations[16] with norm-conserving pseudopotentials using QUANTUM ESPRESSO[17] package. The geometry of the 2H-GeC with 4 atoms unit cell is optimized until the forces on all atoms are less than 10$^{-5}$ eV Å$^{-1}$ and the energy difference is converged to 10$^{-12}$ Ry. A plane-wave cutoff energy of 80 Ry was used. Monkhorst-Pack[18] $k$-point mesh of 12 x 12 x 8 is used during the variable cell optimization. Optimized 2H-GeC structure with lattice constants of a=3.188 Å and c/a=1.646 is shown in Fig. 1 which are in good agreement with the previous study[13]. Elastic constants were computed using QUANTUM ESPRESSO thermo_pw package and Voigt-Reuss-Hill approximation[19] is used to calculate the bulk modulus, shear modulus(G) and Young's Modulus(E). Lattice thermal conductivity is calculated by deriving the most important ingredients, namely, the harmonic and anharmonic interatomic force interactions from density-functional theory and using them with an exact solution of the phonon Boltzmann transport equation (PBTE)[20-22]. Thermal conductivity ($k$) in the single mode relaxation time (SMRT) approximation[23] (usually the first iteration in solution of PBTE) is given by,

$$k_\alpha = \frac{\hbar^2}{N\Omega k_b T^2} \sum_\lambda v_{\alpha\lambda}^2 \omega_\lambda^2 \bar{n}_\lambda (\bar{n}_\lambda + 1)\tau_\lambda \qquad (1)$$

where, $\alpha, \hbar, N, \Omega, k_b, T$, are the cartesian direction, Planck constant, size of the q mesh, unit cell volume, Boltzmann constant, and absolute temperature respectively. $\lambda$ represents the vibrational mode ($qj$) ($q$ is the wave vector and $j$ represent phonon polarization). $\omega_\lambda, \bar{n}_\lambda$, and $v_{\alpha\lambda}$ ($= \partial\omega_\lambda/\partial q$) are the phonon frequency, equilibrium Bose-Einstein population and group velocity along cartesian direction $\alpha$, respectively of a phonon mode $\lambda$. $\omega_\lambda, \bar{n}_\lambda$, and $c_{\alpha\lambda}$ are derived from the knowledge of phonon dispersion computed using 2$^{nd}$ order IFCs. $\tau_\lambda$ is the phonon and is computed using the following equation,

$$\frac{1}{\tau_\lambda} = \pi \sum_{\lambda'\lambda''} |V_3(-\lambda, \lambda', \lambda'')|^2 \times [2(n_{\lambda'} - n_{\lambda''})\delta(\omega(\lambda) + \omega(\lambda') - \omega(\lambda'')) + (1 + n_{\lambda'} + n_{\lambda''})\delta(\omega(\lambda) - \omega(\lambda') - \omega(\lambda''))] \qquad (2)$$

where, $\frac{1}{\tau_\lambda}$ is the anharmonic scattering rate due to intrinsic three phonon interactions and $V_3(-\lambda, \lambda', \lambda'')$ are the three-phonon coupling matrix elements computed using both harmonic and anharmonic interatomic force constants. Dynamical matrix and harmonic force constants were calculated using 8 x 8 x 6 q-grid. 4 x 4 x 3 q-points were used to compute the anharmonic force constants using QUANTUM ESPRESSO D3Q[20, 22, 24] package. Acoustic sum rules were imposed on both harmonic and anharmonic interatomic force constants. Phonon linewidth and lattice thermal conductivity were calculated iteratively using QUANTUM ESPRESSO thermal2 code with 30 x 30 x 24 q -mesh and 0.05 cm$^{-1}$ smearing until the $\Delta k$ values are converged to 1.0e$^{-5}$. Casimir scattering[25] is imposed for length dependence thermal conductivity calculations. Thermal conductivity for naturally occurring 2H-GeC was computed by introducing phonon scattering arising out of mass-disorder due to random distribution of isotopes[26] of Carbon and Germanium throughout the crystal.

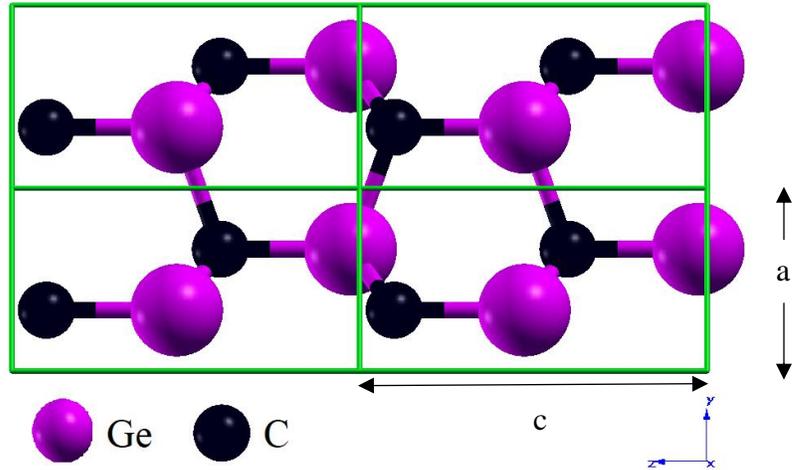

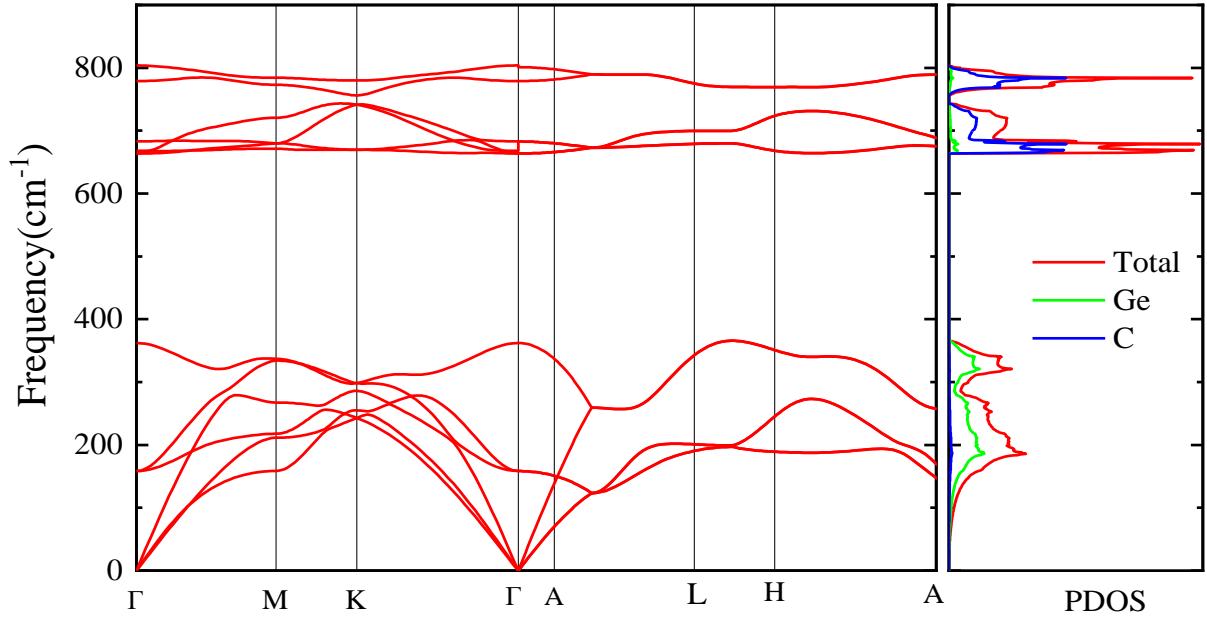

Figure 1: Phonon dispersion and phonon density of state for the 2H-GeC with lattice constants a=3.188 Å and c/a=1.646

Phonon dispersion and phonon density of states for the 2H-GeC with its equilibrium lattice constants a=3.188 Å and c/a=1.646 is shown in Fig 2. Positive phonons frequencies of all the phonon branches indicate the dynamical stability of 2H-GeC. To validate the mechanical stability of the system, we calculated elastic constants to check the Born stability criteria[27]. Elastic constants of 2H-GeC are listed in Table 1 and the values are in good agreements with the previously reported value[13]. The calculated elastic constants satisfied the Born stability criteria of $C_{66}=(C_{11}-C_{12})/2$, $C_{11} > C_{12}$, $C_{33}(C_{11}+C_{12}) > 2(C_{13})$, $C_{44} > 0$, $C_{66} > 0$, and hence the system is mechanically stable. Bulk modulus(B), Young modulus(E) and Shear modulus based on Voigt-Reuss-Hill approximation are also listed in Table 1. The computed values are slightly lower than the hexagonal diamond. These constants indicate the mechanical and dynamical stability of 2H-GeC.

**Table 1: Elastic constants of 2H-GeC**

| Material | C11 | C33 | C44 | C66 | C12 | C13 | Bulk Modulus(B) | Young modulus(E) | Shear Modulus(G) |
|---|---|---|---|---|---|---|---|---|---|
| 2H-GeC | 440.6 | 488.13 | 136.6 | 181 | 78.5 | 36.5 | 185.8 | 389 | 169 |
| 2H-SiC | 522.5 | 557.7 | 156.1 | 214.9 | 92.64 | 43.6 | 218 | 453.2 | 196.4 |

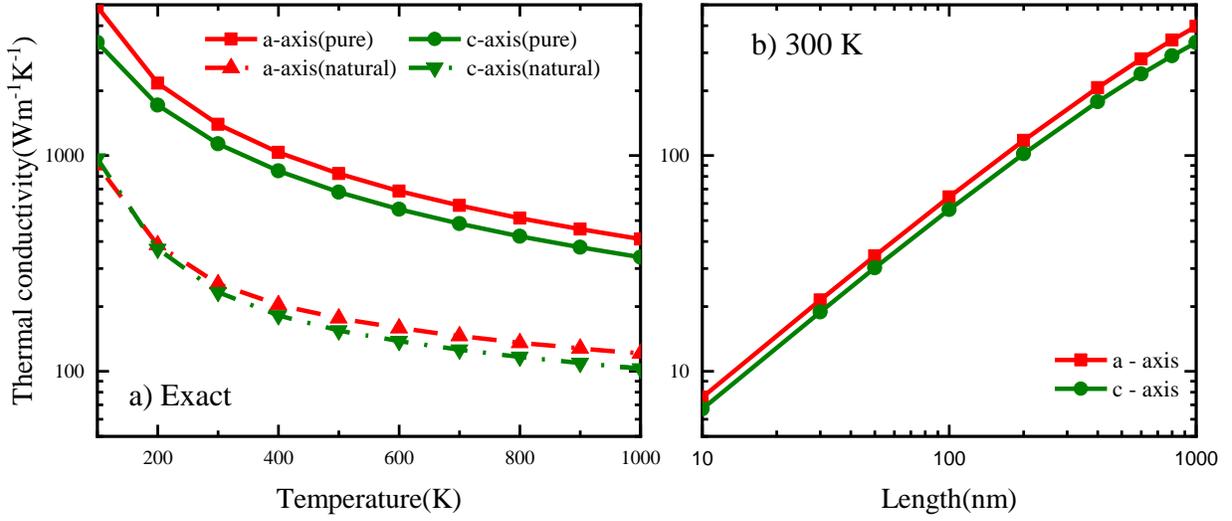

Figure 3a: Temperature dependent lattice thermal conductivity of the isotopically pure and natural 2H-GeC along a and c axis. 3b) Length dependent lattice thermal conductivity of 2H-GeC along a and c axis.

Temperature dependence thermal conductivity($k$) of the pure (solid lines) and naturally (dotted lines) occurring 2H-GeC., obtained by solving the phonon Boltzmann transport equation exactly, is shown along a- and c-axis in Fig 3a. At 300 K, thermal conductivity of pure and naturally occurring 2H-GeC is 1396 Wm$^{-1}$K$^{-1}$(1135 Wm$^{-1}$K$^{-1}$) and 255 Wm$^{-1}$K$^{-1}$(233 Wm$^{-1}$K$^{-1}$) respectively along the a-axis (c-axis). $k$ of naturally occurring 2H-GeC is just 18.2% (20.52%) of the pure 2H-GeC along a-axis which is in good agreement with the cubic germanium carbide(c-GeC)[12]. This is due to the large natural isotopes[26] of germanium (20.57% $^{69.924}$Ge, 27.45% $^{71.922}$Ge, 7.75% $^{72.923}$Ge, 36.5% $^{73.921}$Ge and 7.73% $^{75.921}$Ge) and carbon( 98.93% $^{12}$C and 1.07% 13.003C).

This ultra-high thermal conductivity of 2H-GeC is mainly attributed to high phonon frequencies ($\omega_\lambda$) and phonon group velocities ($= \partial\omega_\lambda/\partial q$) of both acoustic and optical phonons(Shown in Fig.4a) as well as the large phonon bandgap(~335 cm$^{-1}$) between the acoustic and high frequency optical phonons. These high frequencies are due to the strong bonds between the germanium and carbon atoms and the light atomic mass of the constituent atoms C and Ge. These strong bonds are verified by the elastic constants presented in Table 1 with 2H-SiC[28, 29] which are in good agreement with the previous study. Elastic constants of 2H-GeC are slightly lower than the 2H-SiC. This is due to the strong covalent bond between germanium and carbon atoms. Interestingly, we can observe that, optical phonon has a significant phonon group velocity (Fig 4a) and phonon lifetime (Fig 4b). To elucidate its contributions to overall thermal conductivity, we computed the mode dependent thermal conductivity of transverse acoustic ($TA_1$ and $TA_2$), longitudinal acoustic (LA) and optical phonon modes and is shown in Fig 5. We can observe that, at 300 K, optical phonons contribute 621 $Wm^{-1}K^{-1}$(398 $Wm^{-1}K^{-1}$) to its overall thermal conductivity along a-axis(c-axis). This is approximately 46% (45%) to its overall thermal conductivity and is higher than both transverse and longitudinal acoustic phonon modes. This is due to the high phonon group velocities and phonon lifetimes of optical phonons comparable to the acoustic phonons.

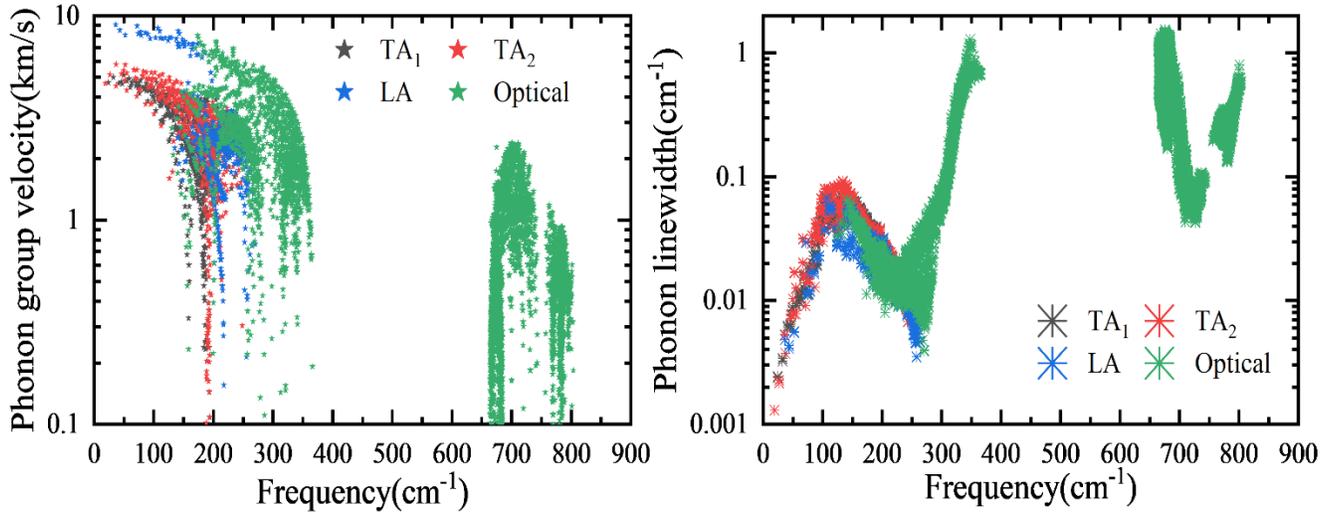

Figure 4a) Phonon group velocity and b) phonon linewidth (inverse of lifetime) for a hexagonal germanium carbide(2H-GeC)

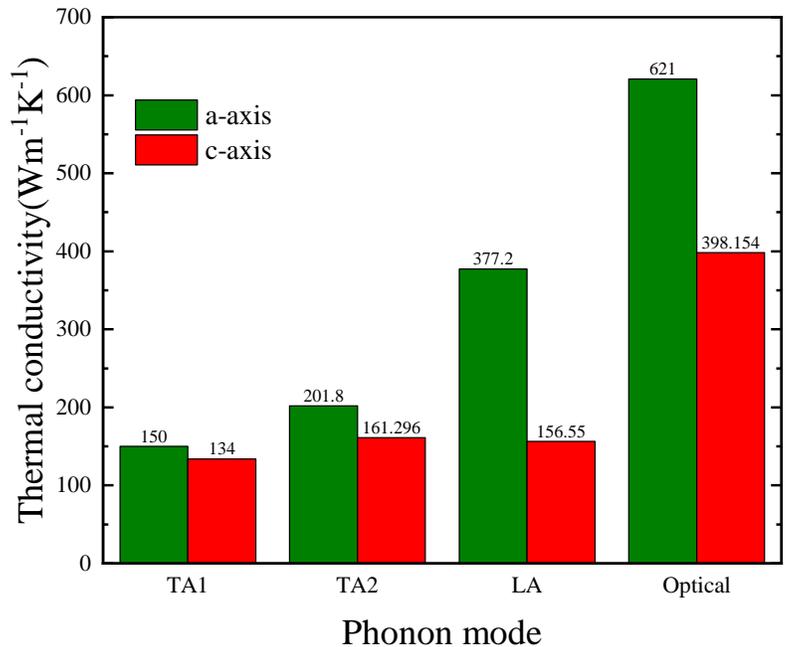

Figure 5: Thermal conductivity contribution from $TA_1$, $TA_2$, LA and optical phonon modes at 300 K.

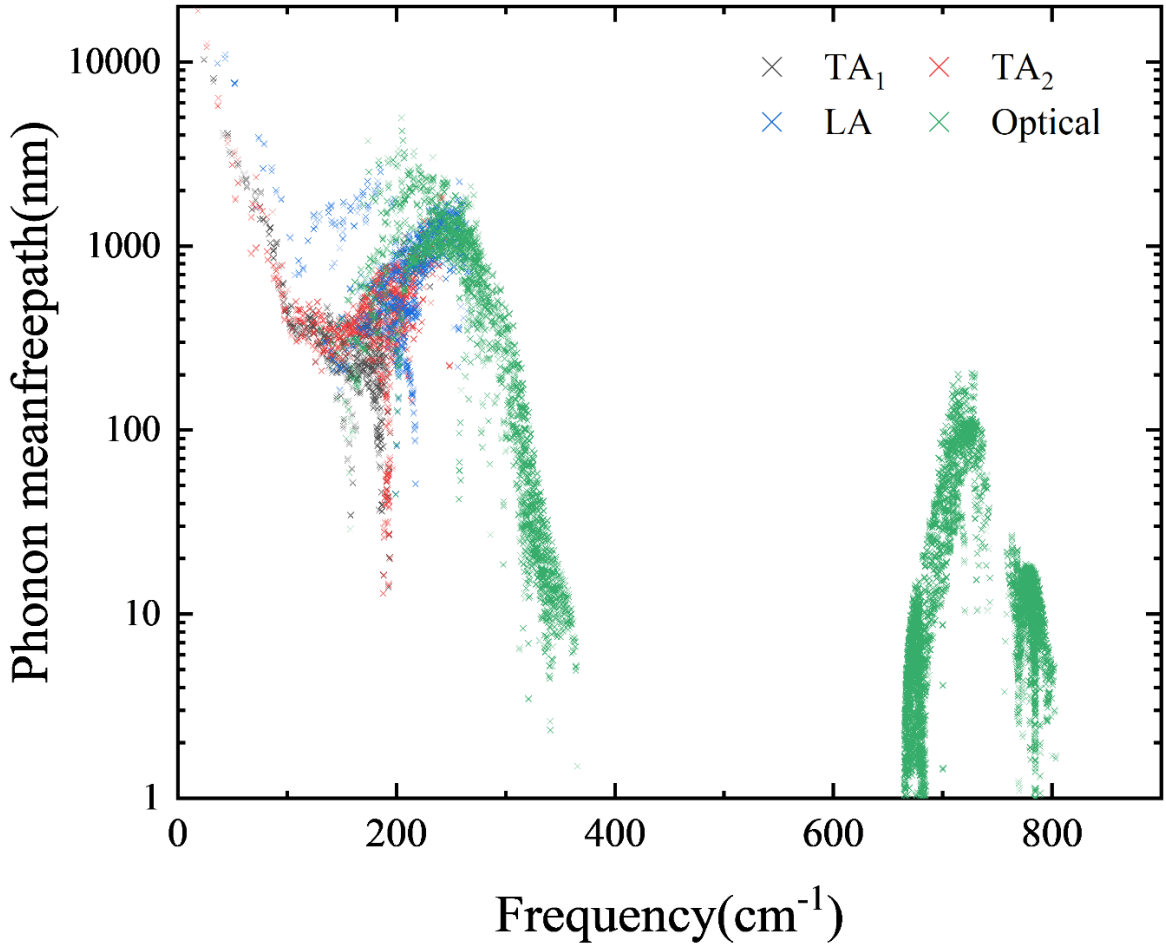

Figure 6: Phonon meanfreepaths(nm) of $TA_1$, $TA_2$, LA and acoustic phonon modes for 2H-GeC at room temperature

We also report the length dependence lattice thermal conductivity between 10 nm and 1000 nm as shown in Fig 3b for the thermal management in nanostructures. At nanoscale length of L=100 nm, room temperature thermal conductivity of 70 $Wm^{-1}K^{-1}$ is 100% higher than that the cubic germanium carbide (Shown in supplementary information). This is due to the large contribution from its optical phonons with phonon meanfreepath in the range of 100nm as shown in Fig 6.

In summary, using first principles calculations, we report the lattice thermal conductivity(*k*) of isotopically pure naturally occurring hexagonal germanium carbide(2H-GeC) by solving the Boltzmann transport equation exactly. At room temperature, we report an ultra-high thermal conductivity of 1350 Wm$^{-1}$K$^{-1}$(1050 Wm$^{-1}$K$^{-1}$) along the a-axis(c-axis) for the pure hexagonal germanium carbide(2H-GeC) which is 130% higher than the hexagonal silicon carbide(2H-SiC). We observed a large reduction (approximately 80%) in *k* due to large isotopes variances of germanium and moderate isotope variance of carbon. We also report the length dependence lattice thermal conductivity for the applications in micro/nanoelectronics. At nanometer length scales of L=100 nm, a high thermal conductivity of 70 Wm$^{-1}$K$^{-1}$ is 100% higher than cubic germanium carbide. These results may lead to potential applications of 2H-GeC in nanoscale thermal management.

**Conflicts of Interest**

There are no conflicts of interest to declare.

**Acknowledgements**

RM and JG acknowledge support from National Science Foundation CAREER award under Award No. #1847129. We also acknowledge OU Supercomputing Center for Education and Research (OSCER) for providing computing resources for this work.

# Supplementary Information

## a) 2H-Silicon carbide

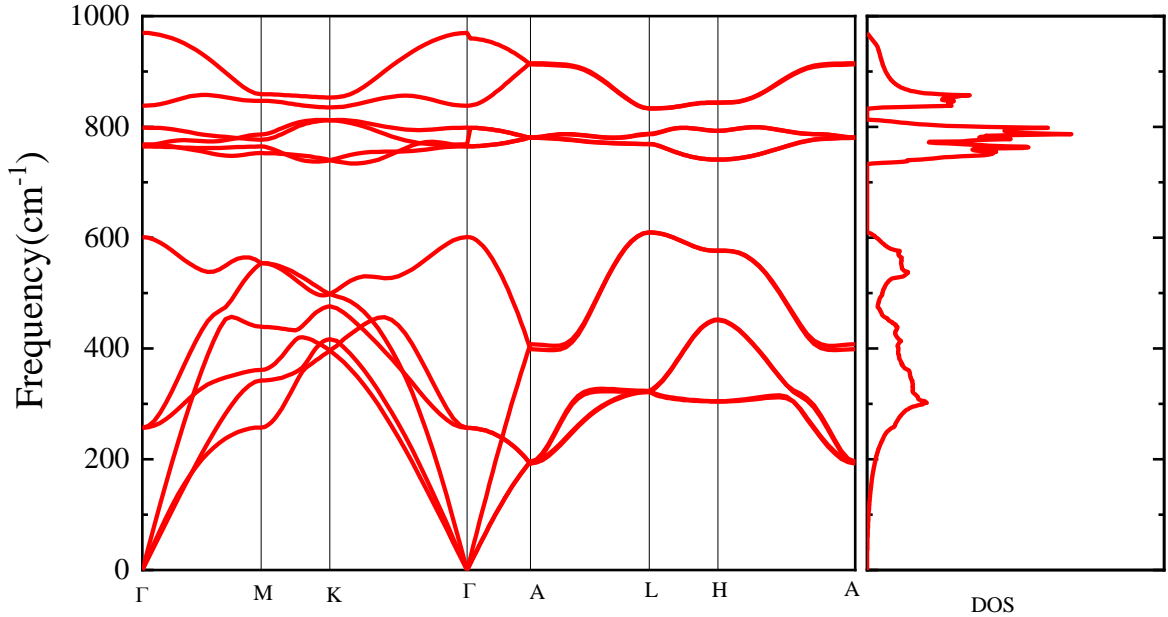

Figure S1: Phonon dispersion for 2H-SiC

We discuss the phonon dispersion and lattice thermal conductivity of 2H-SiC in this supplementary section. A plane-wave cutoff energy of 80 Ry was used. Monkhorst-Pack k-point mesh of 12 x 12 x 8 is used during the variable cell optimization. Lattice parameter for the 2H-SiC is a=3.053 Å and c/a=1.644 which are in good agreement with the previously published values[15]. Dynamical

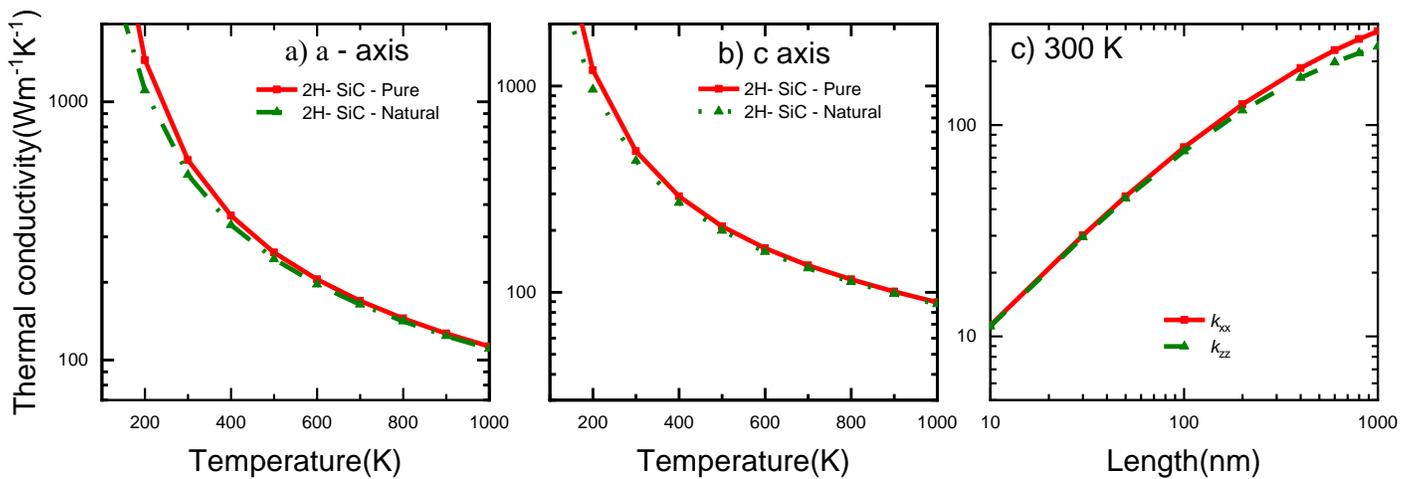

Figure S2: Temperature dependence thermal conductivity of 2H-SiC along a)a-axis and b)c-axis c) Length dependence thermal conductivity of 2H-SiC along a and c-axis

matrix and harmonic force constants were calculated using 8 x 8 x 6 q-grid. 4 x 4 x 3 q-points were used to compute the anharmonic force constants. Thermal conductivity was computed using 30 x 30 x 24 q -mesh and 0.05 cm$^{-1}$ smearing. Temperature dependent thermal conductivity of 2H-SiC is reported in Fig S2a and b along a-axis and c-axis. Length dependent thermal conductivity of 2H-SiC at room temperature is reported in Fig S2c.

### b) Cubic germanium carbide(c_GeC)

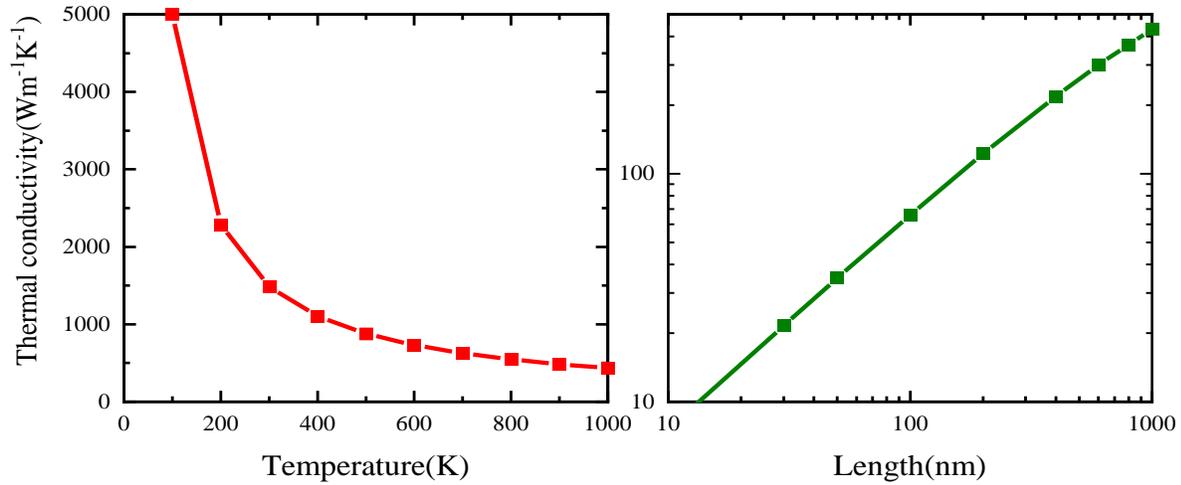

Figure S3a) Temperature and S3b) length dependent lattice thermal conductivity of cubic germanium carbide(c_GeC)

For c-GeC, plane-wave cutoff energy of 70 Ry was used. Monkhorst-Pack k-point mesh of 12 x 12 x 12 is used during the variable cell optimization. Lattice parameter for the c-GeC is a=4.52 Å. Dynamical matrix and harmonic force constants were calculated using 8 x 8 x 8 q-grid. 4 x 4 x 4 q-points were used to compute the anharmonic force constants. Thermal conductivity was computed using 25 x 25 x 25 q -mesh and 0.05 cm$^{-1}$ smearing. Temperature dependent thermal conductivity of c-GeC is reported in Fig S3a. Length dependent thermal conductivity of c-GeC at room temperature is reported in Fig S3b.